\documentclass[conference]{IEEEtran}
\IEEEoverridecommandlockouts
\usepackage{amsmath, amssymb, amsfonts}
\usepackage{graphicx}
\usepackage{textcomp}
\usepackage[utf8]{inputenc}
\usepackage[english]{babel}
\usepackage{framed}
\usepackage{color}
\usepackage{framed}
\usepackage{upgreek}
\usepackage{textcomp}
\usepackage{cite}
\usepackage{bbold}
\usepackage{ifpdf}
\usepackage{times}
\usepackage{layout}
\usepackage{float}
\usepackage{afterpage}
\usepackage{bm}
\usepackage{latexsym}
\usepackage{dblfloatfix}
\usepackage{kantlipsum}
\usepackage{url}
\usepackage[font=scriptsize,caption=false,labelsep=space]{subfig}
\usepackage{booktabs}
\usepackage{multicol}
\usepackage{lipsum}

\usepackage{enumitem}
\usepackage[switch]{lineno}
\usepackage[named]{algo}
\usepackage{algorithm}
\usepackage[noend]{algpseudocode}
\usepackage{seqsplit}

\def\BibTeX{{\rm B\kern-.05em{\sc i\kern-.025em b}\kern-.08em
 T\kern-.1667em\lower.7ex\hbox{E}\kern-.125emX}}
\newcommand{\qqquad}[1][1]{\hspace*{#1em}\ignorespaces}

\usepackage{amsmath}
\usepackage{stackengine}
\usepackage{amsthm}
\usepackage{dsfont}
\usepackage{thmtools}
\usepackage{amssymb}
\usepackage[dvipsnames]{xcolor}

\def\mbu{\mathbf{u}}

\def\mbx{\mathbf{x}}

\def\mbB{\mathbf{B}}

\def\mbD{\mathbf{D}}

\def\mbI{\mathbf{I}}

\def\calK{\mathcal{K}}

\def\calP{\mathcal{P}}

\def\bzero{\boldsymbol{0}}

\newcommand{\realR}[1]{\mathds{R}^{#1}}

\def\rect#1{\mathrm{rect}\left(#1\right)}

\def\Re#1{\operatorname{Re}\left\{#1\right\}}
\def\Im#1{\operatorname{Im}\left(#1\right)}
\def\arg#1{\operatorname{arg}\left\{#1\right\}}

\newtheorem{remark}{Remark}

\def\T{\top}
\def\H{\mathrm{H}}
\def\j{\mathrm{j}}

\begin{document}
\title{Ambiguity Function Shaping in FMCW \\ Automotive Radar
\\
\thanks{* First two authors have equal contributions.}
}
\author{%
\IEEEauthorblockN{Zahra Esmaeilbeig$^{*1\diamond}$, 
Arindam Bose$^{*2\dagger}$, Mojtaba Soltanalian$^{3 \diamond}$}
\IEEEauthorblockA{$^{\diamond}$University of Illinois at Chicago, $^{\dagger}$KMB Telematics Inc.}
\IEEEauthorblockA{\{$^{1}$zesmae2, $^{3}$msol\}@uic.edu, $^{2}$abose@kmb.ac}
}
\maketitle
\begin{abstract}

Frequency-modulated continuous wave (FMCW) radar with inter-chirp coding produces high side-lobes in the Doppler and range dimensions of the radar's ambiguity function. The high side-lobes may cause miss-detection due to masking between targets that are at similar range and have large received power difference, as is often the case in automotive scenarios. In this paper, we develop a novel code optimization method that attenuates the side-lobes of the radar's ambiguity function. In particular, we introduce a framework for designing radar transmit sequences by shaping the radar Ambiguity Function (AF) to a desired structure. The proposed approach suppresses the average amplitude of the AF of the transmitted signal in regions of interest by efficiently tackling a longstanding optimization problem. The optimization criterion is quartic in nature with respect to the radar transmit code. A cyclic iterative algorithm is introduced that recasts the quartic problem as a unimodular quadratic problem (UQP) which can be tackled using power-method-like iterations (PMLI). Our numerical results demonstrate the effectiveness of the proposed algorithm in designing sequences with desired AF which is of great interest to the future generations of automotive radar sensors.
\end{abstract}

\section{Introduction}
In radar signal processing, the range-Doppler response of the transmitted waveform also known as the ambiguity function plays a critical role, as it governs the Doppler and range resolutions of the system and regulates the interference power from unwanted returns at the output of the matched filter to the target signature. To put it another way, the radar designer is faced with the problem of choosing signal waveforms that yield desirable ambiguity functions. What is considered desirable, of course, depends on the operational use of
the radar. While ambiguity function shaping is widely studied in the literature, the topic remains unexplored in the context of automotive radar~\cite{Sussman1962}. 

In this paper, we study the ambiguity function shaping in frequency-modulated continuous wave (FMCW) automotive radar. Shaping radar ambiguity functions has long been considered difficult from a pure design or computational perspective due to the fact that the two-dimensional nature of the ambiguity function implies the number of design constraints would grow much faster than the design variables and that the design objective (to be optimized) has a quartic nature~\cite{aubry2013ambiguity}. In~\cite{aubry2013ambiguity}, a method based on maximum block improvement is devised to tackle the quartic objective in the ambiguity function shaping problem. In a more recent work in~\cite{cui2017local}, an algorithm based on accelerated iterative sequential optimization is proposed to minimize the weighted integrated sidelobes level (WISL) over desired range-Doppler bins of interest. A similar problem is solved by successive application of majorization minimization (MM) and projected gradient descent algorithm (PGD) in~\cite{esmaeili2019unimodular}. This method can be inefficient due to the approximation error in the  MM step. Similar problem is addressed by  convex relaxation in~\cite{bialer2021code}.

In this paper, inspired by the algorithm proposed in~\cite{bose2021mutual,Esmaeilbeig2023mutual} for mutual interference mitigation, we devise a low-complexity algorithm based on power-method-like iterations to minimize the ambiguity function in the range-Doppler bins corresponding to echoes from clutters in the environment. 

The rest of the paper is organized as follows. In the next section, we formulate the ambiguity function shaping problem for FMCW radar. In section~\ref{sec:method}, we propose our algorithm for designing a radar code with the desired ambiguity function. We evaluate our method via numerical experiments in section~\ref{sec:num} and conclude the paper in section~\ref{sec:conclusion}.

Throughout this paper, we use bold lowercase and bold uppercase letters for vectors and matrices, respectively. $\realR{}$ represents the set of real numbers. $(\cdot)^{\top}$ and $(\cdot)^{\H}$ denote the vector/matrix transpose, and the Hermitian transpose, respectively. $\textrm{Diag}(.)$ denotes the diagonalization operator that produces a diagonal matrix with the same diagonal entries as the entries of its vector argument. The $mn$-th element of the matrix $\mbB$ is $\mbB\left[m,n\right]$. The minimum eigenvalue of the matrix $\mbB$ is denoted by $\gamma_{\min}\left(\right)$, respectively. 
The real, imaginary, and angle/phase components of a complex number are $\Re{\cdot}$, $\Im{\cdot}$, and $\arg{\cdot}$, respectively.
Finally, $\delta_{i,j}$ is the extension of Kronecker delta function with $\delta_{i,j}=1$ if $i=j$ and $\delta_{i,j}=0$, otherwise.

\section{Problem Formulation}\label{sec_2}
We start by considering an FMCW automotive radar system whose frequency is swept linearly over a bandwidth
$B$ in a time duration $T_c$. The transmit signal with an intra-pulse code length $N$ can be represented as
\begin{equation}
 s(t)=\sum_{n=1}^{N}x_n u(t-nT_c), \; 0\leq t\leq T_c
\end{equation}
 where $\mbx=[x_1,\ldots,x_N]^{\T} \in \mathbb{C}^N$ is the slow-time sequence and the chirp is 
 \begin{equation}
 u(t)=\frac{1}{\sqrt{T_c}} \exp(\j (2\pi f_c t+\pi K t^2)) \rect{\frac{t}{T_c}},
 \end{equation}
where $K=\frac{B}{T_c}$ is the chirp rate, and
\begin{equation}
\rect{t}=\begin{cases}
1 & 0 \leq t \leq 1, \\
0 & \text{otherwise}.
\end{cases}
\end{equation}
In order to keep constant transmit power over the $N$ chirps, we constrain the code sequence to be unimodular i.e. $|x_n|=1$, for $n=1, \ldots, N$~\cite{bose2022waveform,Uysal}. 

The ambiguity function is defined as \cite{He2011CAF}, \par \noindent \small
\begin{align} \label{eq:af-def}
 &\chi(\tau,\nu) \\ &= \int_{-\infty}^{\infty} s(t) s^*(t-\tau) \exp{(-\j 2\pi \nu(t-\tau))}\; dt \nonumber \\&=\int_0^T \left( \sum_{n=1}^N{x_n u(t-nT_c)} \right) \left( \sum_{m=1}^N{x_m^* u^*(t-mT_c-\tau)} \right) \nonumber \\ & \qqquad[14] \cdot \exp{(-\j 2\pi \nu(t-\tau))}\; dt \nonumber\\ &=\sum_{m=1}^N\sum_{n=1}^N x_m^* \left( \int_0^T{u(t-nT_c)u^*(t-mT_c-\tau) e^{-\j 2\pi \nu(t-\tau)} dt} \right) x_n. \nonumber
\end{align}\normalsize
where $\tau$ is the time delay and $\nu$ is the Doppler frequency shift.
With an aim to discretize the AF in \eqref{eq:af-def}, by setting $\tau=kT_c$ for $k=-N+1, \cdots, 0, \cdots, N-1$ and $\nu=\frac{p}{NT_c}$ for $p=-\frac{N}{2}, \cdots, \frac{N}{2}-1$ for even $p$ or $p=-\frac{N-1}{2}, \cdots, \frac{N-1}{2}$ for odd $p$, we can easily obtain,
\begin{align}
 \chi[k, p] &\triangleq \chi(kT_c, \dfrac{p}{NT_c}) \nonumber\\
 &= e^{\j\pi \frac{p}{N}} \text{sinc}\left(\pi\dfrac{p}{N}\right) \sum_{n=1}^N{x_n x_{n-k}^* e^{-\j\pi(n-k)p/N}}.
\end{align}
where $\text{sinc}(x) = \sin(x)/x$.
We assume the target under study is moving with low speed i.e. $|\nu| \ll 1/T_c$. Therefore, it is safe to confine our attention to values of $|p| \ll N$ in which case $\text{sinc}\left(\pi\dfrac{p}{N}\right) \approx 1$ and thus the discrete-AF can be defined as,
\begin{align}
 r[k,p] \triangleq \sum_{n=1}^N {x_nx_{n-k}^* e^{-\j2\pi\frac{(n-k)p}{N}}}.
\end{align}
for $k=-N+1, \cdots, 0, \cdots, N-1$ and $p=-\frac{N}{2}, \cdots, \frac{N}{2}-1$ for even $p$ or $p=-\frac{N-1}{2}, \cdots, \frac{N-1}{2}$ for odd $p$. 
In the next section, we will primarily be focused on designing the sequence $\{x_n\}^N_{n=1}$ so as to minimize the sidelobes of the discrete-AF in a certain region.

\section{Proposed Method} \label{sec:method}
The goal herein is to suppress the energy of the discrete-AF in a region of interest defined by the index sets $\mathcal{K}, \mathcal{P}$ for delay and Doppler shift, respectively, by minimizing the criterion:
\begin{align}\label{eq:C}
 C=\sum_{k\in \mathcal{K}}\sum_{p\in\mathcal{P}}\left|r[k,p]\right|^2.
\end{align}
In particular, the AF shaping problem that we are interested in is 
\begin{align}\label{eq:19}
\mathcal{M}_1 : ~\underset{\mbx}{\textrm{minimize}} &\quad C\nonumber\\
\text{s.t.}~&\mathbf{x} ~\text{is unimodular}. 
\end{align} Note that the discrete-AF $r[k,p]$ can be reformulated as
\begin{align}
 r[k,p]=\mathbf{x}^\H \mathbf{D}_{p} \mathbf{J}_k \mathbf{x},
\end{align}
where 
\begin{align}
 \mathbf{D}_{p} &=\text{Diag}\left(\left[e^{-\j2\pi\frac{p}{N}}, \cdots, e^{-\j2\pi\frac{(N-1)p}{N}},e^{-\j2\pi\frac{Np}{N}}\right]\right),
\end{align}
and
\begin{equation}
 \mathbf{J}_k = \mathbf{J}_{-k}^\H= \begin{bmatrix} \bzero & \mbI_{N-k} \\ \mbI_k & \bzero \end{bmatrix}
\end{equation}
is the shift matrix that performs the shifting of the vector being multiplied by $k$ lags.
Therefore the problem in~\eqref{eq:C} can be recast as,
\begin{align}
 C&=\sum_{k\in \mathcal{K}}\sum_{p\in\mathcal{P}}|\mathbf{x}^\H \mathbf{D}_{p} \mathbf{J}_k \mathbf{x}|^2 \nonumber\\
 &=\sum_{k\in \mathcal{K}}\sum_{p\in\mathcal{P}}|\mathbf{x}^\H \mathbf{A}_{k,p} \mathbf{x}|^2
\end{align}
where $\mathbf{A}_{k,p} = \mathbf{D}_{p}\mathbf{J}_k$.
Interestingly, as one can observe, $C$ is quartic with respect to $\mathbf{x}$ making $\mathcal{M}_1$ in \eqref{eq:19} a non-convex problem.
In order to recast the problem in a quadratic form, let
\begin{align}\label{eq:14}
 \mathbf{A}_{k,p}^r &\triangleq \frac{1}{2}(\mathbf{A}_{k,p} + \mathbf{A}_{k,p}^\H),\nonumber\\ \mathbf{A}_{k,p}^i &\triangleq \frac{1}{2}(\mathbf{A}_{k,p} - \mathbf{A}_{k,p}^\H)
\end{align}
and note that
\begin{enumerate}
 \item Matrices $\mathbf{A}_{k,p}^r$ and $\j\mathbf{A}_{k,p}^i$ are Hermitian and skew-Hermitian matrices, respectively~\cite{Hu2017locating}.
 \item For any generic vector $\mathbf{z}$,
 \begin{align} \label{eq:15}
 \mathbf{z}^\H\mathbf{A}_{k,p}\mathbf{z} = \mathbf{z}^\H\mathbf{A}_{k,p}^r\mathbf{z} + \mathbf{z}^\H\mathbf{A}_{k,p}^i\mathbf{z}
 \end{align}
 where
 \begin{equation} \label{eq:16}
 \mathbf{z}^{\H} \mathbf{A}_{k,p}^r \mathbf{z} \in \realR{} \qquad \text{and} \quad
 \j\mathbf{z}^{\H} \mathbf{A}_{k,p}^i \mathbf{z} \in \realR{} . 
 \end{equation}
 In particular, it follows from~\eqref{eq:16} that
 \begin{align} \label{eq:17}
 |\mathbf{z}^\H\mathbf{A}_{k,p}\mathbf{z}|^2 = |\mathbf{z}^\H\mathbf{A}_{k,p}^r\mathbf{z}|^2 + |\mathbf{z}^\H\j\mathbf{A}_{k,p}^i\mathbf{z}|^2.
 \end{align}
\end{enumerate}
Hence we can write,
\begin{align}\label{eq:18}
 \sum_{k,p}|\mathbf{x}^\H \mathbf{A}_{k,p} \mathbf{x}|^2 = &\sum_{k,p}|\mathbf{x}^\H \mathbf{A}_{k,p}^r \mathbf{x}|^2 + |\mathbf{x}^\H j\mathbf{A}_{k,p}^i \mathbf{x}|^2 \\ 
 = &\sum_{k,p} |\mathbf{x}^\H (\mathbf{A}_{k,p}^r + \zeta\mathbf{I}_N) \mathbf{x} - \zeta N|^2 \nonumber\\&+ |\mathbf{x}^\H (\j\mathbf{A}_{k,p}^i + \zeta\mathbf{I}_N) \mathbf{x} - \zeta N|^2 \nonumber\\
 = &\sum_{k,p}|\mathbf{x}^\H \mathbf{\tilde A}_{k,p}^r \mathbf{x} - \zeta N|^2 + |\mathbf{x}^\H \mathbf{\tilde A}_{k,p}^i \mathbf{x} - \zeta N|^2 \nonumber
\end{align}
where
\begin{align}
 \mathbf{\tilde A}_{k,p}^r &= \mathbf{A}_{k,p}^r + \zeta\mathbf{I}_N, \label{eq:diag1}\\
 \mathbf{\tilde A}_{k,p}^i &= \j\mathbf{A}_{k,p}^i + \zeta\mathbf{I}_N \label{eq:diag2}
\end{align}
and $\zeta \in \realR{}$ is chosen such that \begin{align}
 \zeta > - \min \left( \bigcup_{k,p}\left\lbrace \gamma_{\min}\left(\mathbf{A}_{k,p}^r\right), \gamma_{\min}\left(\j\mathbf{A}_{k,p}^i
 \right) \right\rbrace\right).
\end{align}
The modification in~\eqref{eq:diag1}-\eqref{eq:diag2} is  known as  diagonal loading and it ensures the positive definiteness of 
$\{\mathbf{\tilde A}_{k,p}^r\}$ and 
$\{\mathbf{\tilde A}_{k,p}^i\}$. The objective~\eqref{eq:18} is still quartic w.r.t. $\mathbf{x}$. In order to make it quadratic we resort to the equivalence properties of Hermitian square roots.
\begin{remark}
For the positive definite matrix $\mathbf{\tilde A}_{k,p}^r$, $\mathbf{x}^\H \mathbf{\tilde A}_{k,p}^r \mathbf{x}$ is close to $\zeta N$, if and only if $(\mathbf{\tilde A}_{k,p}^r)^{1/2}\mbx$ is close to $\sqrt{\zeta N} \mathbf{u}_{k,p}^r$, for a unit-norm vector $\mbu_{k,p}^r$. Similarly, $\mathbf{x}^H \mathbf{\tilde A}_{k,p}^i \mathbf{x}$ is close to $\zeta N$, if and only if $(\mathbf{\tilde A}_{k,p}^i)^{1/2}\mbx$ is close to $\sqrt{\zeta N} \mathbf{u}_{k,p}^i$, for a unit-norm vector $\mbu_{k,p}^r$~\cite{Hu2017locating}.
\end{remark}
According to Remark 1,  $\mathcal{M}_1$  is equivalent to  
\begin{align}\label{eq:p2}
 \mathcal{M}_2 : ~\underset{\mbx,\{\mathbf{u}_{k,p}^r\},\{\mathbf{u}_{k,p}^i\}}{\textrm{minimize}} \quad &\sum_{k,p} \left\lbrace \left\|(\mathbf{\tilde A}_{k,p}^r)^{1/2}\mathbf{x} - \sqrt{\zeta N} \mathbf{u}_{k,p}^r\right\|_2^2 \right.\nonumber \\
&\left.\qquad+ \left\|(\mathbf{\tilde A}_{k,p}^i)^{1/2}\mathbf{x} - \sqrt{\zeta N}\mathbf{u}_{k,p}^i \right\|_2^2 \right\rbrace \nonumber\\
\text{s.t.}~\mathbf{x}&~\text{is unimodular}, \nonumber\\ 
\|\mathbf{u}_{k,p}^r\|_2&=\|\mathbf{u}_{k,p}^i\|_2 = 1 ~\text{for all}~k\in \mathcal{K},p\in \mathcal{P},
\end{align}
which is quadratic w.r.t. $\mbx,\{\mathbf{u}_{k,p}^r\}$ and $\{\mathbf{u}_{k,p}^i\}$ and equivalent to $\mathcal{M}_1$ in~\eqref{eq:19}. In the following, we follow a cyclic optimization approach to tackle the problem~\eqref{eq:19} in an alternating manner over $\mbx$, $\{\mathbf{u}_{k,p}^r\}$ and $\{\mathbf{u}_{k,p}^i\}$.
\subsection{Optimization w.r.t. $\mathbf{x}$}
The objective function in $\mathcal{M}_2$ is recast as
\begin{align}\label{eq:C_x}
 C_{\mathbf{x}} &= \mathbf{x}^H \left(\sum_{k,p}\left (\mathbf{\tilde A}_{k,p}^r + \mathbf{\tilde A}_{k,p}^i \right) \right) \mathbf{x} \nonumber\\ 
 &~~- 2\sqrt{\zeta N}\Re{\mathbf{x}^H \sum_{k,p} (\mathbf{\tilde A}_{k,p}^r)^{H/2}\mathbf{u}_{k,p}^{r}} \nonumber\\
 &~~- 2\sqrt{\zeta N}\Re{\mathbf{x}^H \sum_{k,p} (\mathbf{\tilde A}_{k,p}^i)^{H/2}\mathbf{u}_{k,p}^{i}} + \text{const.}
\end{align}
Or simply, 
\begin{align}
 C_{\mathbf{x}} = \mathbf{x}^H \mathbf{R}\mathbf{x} + 2\Re{\mathbf{x}^H \mathbf{s}_{\mathbf{x}}} + \text{const.}
\end{align}
where
\begin{align}
 \mathbf{R}= \sum_{k,p}\left (\mathbf{\tilde A}_{k,p}^r + \mathbf{\tilde A}_{k,p}^i \right)
\end{align}
and
\begin{align}\label{eq:20}
 \mathbf{s}_{\mathbf{x}} = - \sqrt{\zeta N} \sum_{k,p}\left((\mathbf{\tilde A}_{k,p}^r)^{H/2}\mathbf{u}_{k,p}^{r} + (\mathbf{\tilde A}_{k,p}^i)^{H/2}\mathbf{u}_{k,p}^{i}\right)
\end{align} 
By dropping the constant term, the objective function can be reformulated as,
\begin{align}
 C_{\mathbf{x}} &= \mathbf{x}^H \mathbf{R}\mathbf{x} + 2\Re{\mathbf{x}^H \mathbf{s}_{\mathbf{x}}} \nonumber\\
 &=\begin{bmatrix} \mathbf{x} \\ 1\end{bmatrix}^H \begin{bmatrix} \mathbf{R} & \mathbf{s}_{\mathbf{x}} \\ \mathbf{s}_{\mathbf{x}}^H & 0\end{bmatrix} \begin{bmatrix} \mathbf{x} \\ 1 \end{bmatrix} \nonumber\\ 
 &= \bar{\mathbf{x}}^H \mathbf{B}_{\mathbf{x}} \bar{\mathbf{x}}
\end{align}
Hence, $\mathcal{M}_2$ w.r.t $\mathbf{x}$ is equivalent to 
\begin{align}
 \underset{\bar{\mathbf{x}}} {\textrm{minimize}}&~~\bar{\mathbf{x}}^H \mathbf{B}_{\mathbf{x}} \bar{\mathbf{x}} \nonumber\\ \text{s.t.}&~~|x_n|=1, ~~n=1,\cdots,N, \nonumber\\ & \bar{\mathbf{x}} = \begin{bmatrix} \mathbf{x} \\ 1 \end{bmatrix}.
\end{align}
We  perform  diagonal loading on $\mbB_{\mbx}$ to obtain the equivalent problem
\begin{align}\label{eq:21}
 \underset{\bar{\mathbf{x}}} {\textrm{maximize}} &~~\bar{\mathbf{x}}^\H \mathbf{D}_{\mathbf{x}} \bar{\mathbf{x}} \nonumber\\ \text{s.t.}&~~|x_n|=1, ~~n=1,\cdots,N, \nonumber\\ & \bar{\mathbf{x}} = \begin{bmatrix} \mathbf{x} \\ 1 \end{bmatrix}.
\end{align}
where $\mathbf{D}_{\mathbf{x}} \triangleq \gamma_{\mathbf{x}} I_{(N+1)}-\mathbf{B}_{\mathbf{x}}$, with $\gamma_{\mathbf{x}}$ being larger than the maximum eigenvalue of $\mathbf{B}_{\mathbf{x}}$. 
The above problem is called unimodular quadratic programming (UQP) and the power-method-like iterations,
\begin{align}\label{eq:xx}
 \mathbf{x}^{(t,s+1)} = \exp\left\lbrace \j\arg{\begin{bmatrix} \mathbf{I}_{N\times N} \\ \mathbf{0}_{1\times N} \end{bmatrix}^T \mathbf{D}_{\mathbf{x}} \bar{\mathbf{x}}^{(t,s)}}\right\rbrace
\end{align}
introduced in~\cite{Soltanalian2014UQP} leads to a monotonically decreasing objective value for UQP. The iterations can be initialized with the latest design of $\mathbf{x}$ denoted by $\mathbf{x}^{(t,0)}$, where $t$ denotes the iteration number as we see later in Algorithm 1.
\subsection{Optimization w.r.t. $\{\mathbf{u}^r_{k,p}\}$ and $\{\mathbf{u}^i_{k,p}\}$}
Using~\eqref{eq:C_x}, the problem $\mathcal{M}_2$ w.r.t $\mathbf{u}^r_{k,p}$ is equivalent to
\begin{align}\label{eq:p2}
~\underset{\mathbf{u}_{k,p}^r}{\textrm{minimize}} &~~\Re{\mathbf{x}^H (\mathbf{\tilde A}_{k,p}^r)^{H/2}\mathbf{u}_{k,p}^{r}} \nonumber \\
 \text{s.t.}&~~ \|\mathbf{u}_{k,p}^r\|_2 = 1.
\end{align}
Therefore, we have the closed-form solution for $\mathbf{u}^r_{k,p}$ as 
\begin{align}\label{eq:u_r}
 \widehat{\mathbf{u}}_{k,p}^{r(t)} &= \frac{(\mathbf{\tilde A}_{k,p}^r)^{1/2}\mathbf{x}}{\|(\mathbf{\tilde A}_{k,p}^r)^{1/2}\mathbf{x}\|_2},
\end{align}
where $t$ is the iteration number as used in Algorithm 1. A similar  closed-form solution works mutatis mutandis
for $\mathbf{u}^i_{k,p}$  as 
\begin{align}\label{eq:u_i}
 \widehat{\mathbf{u}}_{k,p}^{i(t)} &= \frac{(\mathbf{\tilde A}_{k,p}^i)^{1/2}\mathbf{x}}{\|(\mathbf{\tilde A}_{k,p}^i)^{1/2}\mathbf{x}\|_2}.
\end{align}
At each cycle of the algorithm, we  compute~\eqref{eq:u_r}-\eqref{eq:u_i} corresponding to each $k \in \mathcal{K}$ and  $p \in \mathcal{P}$. 
The final algorithm consisting of iterations over~\eqref{eq:xx} and~\eqref{eq:u_r}-\eqref{eq:u_i}  is summarized in Algorithm 1. The number of outer  iterations , $\Gamma_1$, in the algorithm is chosen such that $|(C^{(t+1)}-C^{(t)})/C^{(t)}|\leq \epsilon$, where $C^{(t)}$ is the objective value introduced in~\eqref{eq:C}, is satisfied at the final iteration. Similarly, the number of inner iterations $\Gamma_2$ is chosen such that the power method like iterations in~\eqref{eq:xx} for updating  $\mbx$  converges in terms of changes in objective value.
\begin{algorithm}[H]
\caption{Radar code design for shaping the ambiguity function }
 \label{algorithm_1}
 \begin{algorithmic}[1]
 \Statex \textbf{Input:} Index sets $\calK$ and $\calP$, $\mbx^{(0,0)}$, $\mbu_{k,p}^{r(0)}$, $\mbu_{k,p}^{i(0)}$ for $k \in \calK$ and $p \in \calP, \Gamma_{1}, \Gamma_{2}$.
 \Statex \textbf{Output:} $\mbx$
 
 \For{$t=0:\Gamma_{1}-1$} 
 \For{$s=0:\Gamma_{2}-1$} 
 \State Update $\mbD_x$ by plugging in $\widehat{\mathbf{u}}_{k,p}^{r(t)}$ and $\widehat{\mathbf{u}}_{k,p}^{i(t)}$ in ~\eqref{eq:20}-~\eqref{eq:21}.
 
 \State$\mbx^{(t,s+1)} \leftarrow \exp\left\lbrace \j\arg{\begin{bmatrix} \mathbf{I}_{N\times N} \\ \mathbf{0}_{1\times N} \end{bmatrix}^T \mathbf{D}_{\mathbf{x}} \bar{\mathbf{x}}^{(t,s)}}\right\rbrace$
 \EndFor
 \State$\widehat{\mathbf{u}}_{k,p}^{r(t+1)} \gets \frac{(\mathbf{\tilde A}_{k,p}^r)^{1/2}\mbx^{(t,s)}}{\|(\mathbf{\tilde A}_{l,p}^r)^{1/2}\mbx^{(t,s)}\|_2},$   $k \in \mathcal{K}$ and  $p \in \mathcal{P}.$\\ 
 \State$\widehat{\mathbf{u}}_{k,p}^{i(t+1)} \gets \frac{(\mathbf{\tilde A}_{k,p}^i)^{1/2}\mbx^{(t,s)}}{\|(\mathbf{\tilde A}_{k,p}^i)^{1/2}\mbx^{(t,s)}\|_2}$, $k \in \mathcal{K}$ and  $p \in \mathcal{P}.$
 
 \EndFor
 \State \Return $\mbx \gets \mbx^{(\Gamma_{1},\Gamma_{2})}$
 \end{algorithmic}
\end{algorithm}
\begin{figure}[t]
\centering
	\includegraphics[width=1\columnwidth]{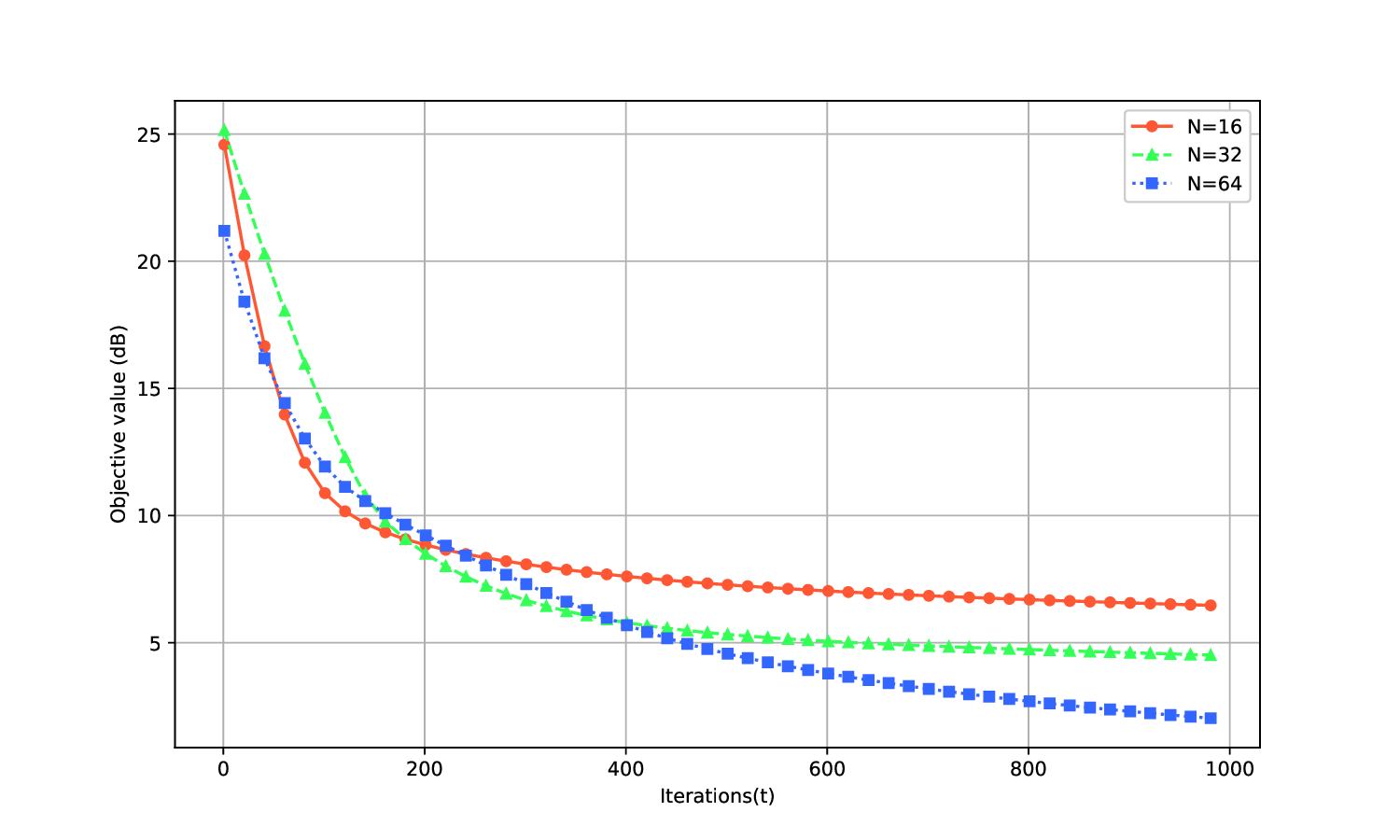}
	\caption{The objective value in ~\eqref{eq:C} versus the iterations of Algorithm 1}
	\label{fig::result1}
\end{figure}
\section{Numerical Experiments}\label{sec:num}
\begin{figure}
\footnotesize
\stackunder[5pt]{\includegraphics[width=1\columnwidth]{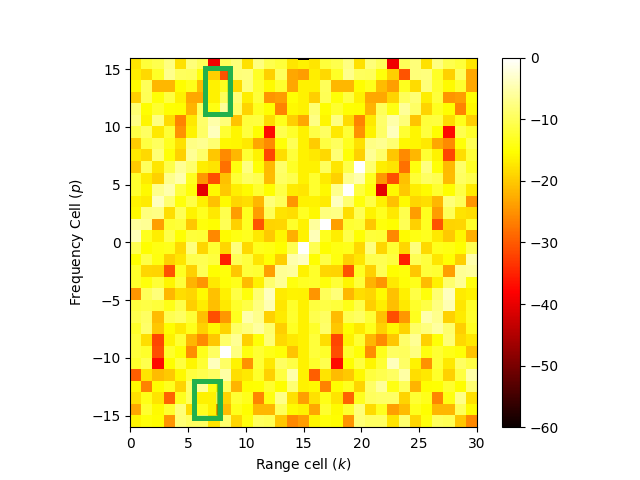}}{(a)}%
\hspace{1cm}%
\stackunder[5pt]{\includegraphics[width=1\columnwidth]{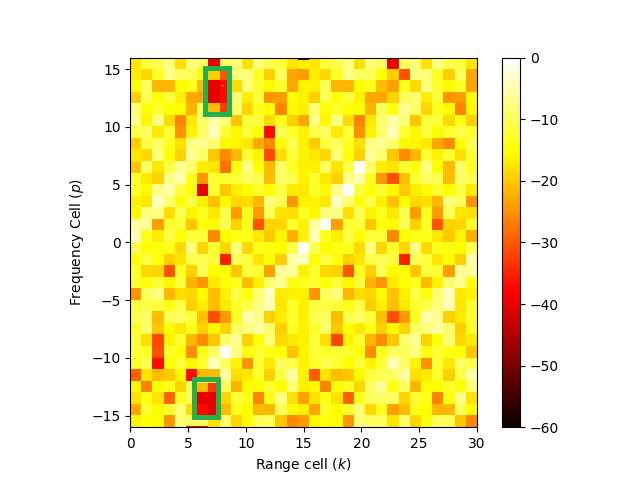}}{(b)}
\caption{Ambiguity function, in dB, of (a) the initial random code and (b) the synthesized FMCW code with $N=16$ and in green the assumed regions of interest. } \label{fig-AF1}
\end{figure}

In this section, we will examine the capability of Algorithm 1 which has been proposed to design a radar phase code that has an ambiguity function with the desired shape. The region of interest is defined by the sets $\mathcal{K}$ and $\mathcal{P}$ as
\begin{align}
\mathcal{K}&=\{5,6,7\} \qquad \text{and} \nonumber\\
\mathcal{P}&=\{-15,-14,-13,11,12,13,14\}.
\end{align}
A random phase-code unimodular sequence of length $N=31$  is generated as the starting sequence for the algorithm. Moreover, we execute the UQP subroutine for $\Gamma_2=500$ times and allow for at most $\Gamma_1=10^3$ runs of the outer iterations. As illustrated in Fig.~\ref{fig-AF1}, the radar code synthesized by algorithm 1 has the desired ambiguity function values in the chosen bins corresponding to interference. 

\section{Summary}\label{sec:conclusion}
In this paper, we addressed the unimodular radar code design for FMCW radar ambiguity function shaping. We devised the radar codes by minimizing a criterion obtained from the absolute value of the ambiguity function in the regions of interest and we addressed the quartic optimization problem using the PMLI iterations. Numerical experiments were conducted to demonstrate the efficacy of the proposed method in shaping the ambiguity function.
\bibliographystyle{IEEEtran}
\bibliography{refs}

\begin{thebibliography}{10}
\providecommand{\url}[1]{#1}
\csname url@samestyle\endcsname
\providecommand{\newblock}{\relax}
\providecommand{\bibinfo}[2]{#2}
\providecommand{\BIBentrySTDinterwordspacing}{\spaceskip=0pt\relax}
\providecommand{\BIBentryALTinterwordstretchfactor}{4}
\providecommand{\BIBentryALTinterwordspacing}{\spaceskip=\fontdimen2\font plus
\BIBentryALTinterwordstretchfactor\fontdimen3\font minus
  \fontdimen4\font\relax}
\providecommand{\BIBforeignlanguage}[2]{{%
\expandafter\ifx\csname l@#1\endcsname\relax
\typeout{** WARNING: IEEEtran.bst: No hyphenation pattern has been}%
\typeout{** loaded for the language `#1'. Using the pattern for}%
\typeout{** the default language instead.}%
\else
\language=\csname l@#1\endcsname
\fi
#2}}
\providecommand{\BIBdecl}{\relax}
\BIBdecl

\bibitem{Sussman1962}
S.~Sussman, ``Least-square synthesis of radar ambiguity functions,'' \emph{IRE
  Transactions on Information Theory}, vol.~8, no.~3, pp. 246--254, 1962.

\bibitem{aubry2013ambiguity}
A.~Aubry, A.~De~Maio, B.~Jiang, and S.~Zhang, ``Ambiguity function shaping for
  cognitive radar via complex quartic optimization,'' \emph{IEEE Transactions
  on Signal Processing}, vol.~61, no.~22, pp. 5603--5619, 2013.

\bibitem{cui2017local}
G.~Cui, Y.~Fu, X.~Yu, and J.~Li, ``Local ambiguity function shaping via
  unimodular sequence design,'' \emph{IEEE Signal Processing Letters}, vol.~24,
  no.~7, pp. 977--981, 2017.

\bibitem{esmaeili2019unimodular}
H.~Esmaeili-Najafabadi, H.~Leung, and P.~W. Moo, ``Unimodular waveform design
  with desired ambiguity function for cognitive radar,'' \emph{IEEE
  Transactions on Aerospace and Electronic Systems}, vol.~56, no.~3, pp.
  2489--2496, 2019.

\bibitem{bialer2021code}
O.~Bialer, A.~Jonas, and T.~Tirer, ``Code optimization for fast chirp {FMCW}
  automotive {MIMO} radar,'' \emph{IEEE Transactions on Vehicular Technology},
  vol.~70, no.~8, pp. 7582--7593, 2021.

\bibitem{bose2021mutual}
A.~Bose, B.~Tang, M.~Soltanalian, and J.~Li, ``Mutual interference mitigation
  for multiple connected automotive radar systems,'' \emph{IEEE Transactions on
  Vehicular Technology}, vol.~70, no.~10, pp. 11\,062--11\,066, 2021.

\bibitem{Esmaeilbeig2023mutual}
Z.~Esmaeilbeig, A.~Bose, and M.~Soltanalian, ``Mutual interference mitigation
  in {PMCW} automotive radar,'' in \emph{20th European Radar Conference}, 2023,
  pp. 118--121.

\bibitem{bose2022waveform}
A.~Bose, B.~Tang, W.~Huang, M.~Soltanalian, and J.~Li, ``Waveform design for
  mutual interference mitigation in automotive radar,'' \emph{arXiv preprint
  arXiv:2208.04398}, 2022.

\bibitem{Uysal}
F.~Uysal and S.~Orru, ``Phase-coded {FMCW} automotive radar: Application and
  challenges,'' in \emph{IEEE International Radar Conference}, 2020, pp.
  478--482.

\bibitem{He2011CAF}
H.~He, P.~Stoica, and J.~Li, ``On synthesizing cross ambiguity functions,'' in
  \emph{2011 IEEE International Conference on Acoustics, Speech and Signal
  Processing (ICASSP)}, 2011, pp. 3536--3539.

\bibitem{Hu2017locating}
H.~{Hu}, M.~{Soltanalian}, P.~{Stoica}, and X.~{Zhu}, ``Locating the few:
  Sparsity-aware waveform design for active radar,'' \emph{IEEE Transactions on
  Signal Processing}, vol.~65, no.~3, pp. 651--662, 2017.

\bibitem{Soltanalian2014UQP}
M.~Soltanalian and P.~Stoica, ``Designing unimodular codes via quadratic
  optimization,'' \emph{IEEE Transactions on Signal Processing}, vol.~62,
  no.~5, pp. 1221--1234, 2014.

\end{thebibliography}
\end{document}